\documentclass[aps,prl,twocolumn,showpacs]{revtex4}
\usepackage[dvips]{graphicx}
\usepackage{amsmath}
\usepackage[latin1]{inputenc}
\usepackage[english]{babel}

\begin{document}

\title{Free-space lossless state-detection of a single trapped atom}

\author{A. Fuhrmanek, R. Bourgain, Y.R.P. Sortais, and A. Browaeys}

\affiliation{Laboratoire Charles Fabry, Institut d'Optique, CNRS, Univ Paris-Sud,
Campus Polytechnique, 2 avenue Augustin Fresnel, RD 128,
91127 Palaiseau cedex, France}

\date{\today}

\begin{abstract}

We demonstrate the lossless state-selective detection of a single rubidium 87 atom trapped  in an optical tweezer. This detection is analogous to the one used on trapped ions. After preparation in either a dark or bright state, we probe the atom internal state by sending laser light that couples an excited state to the bright state only. The laser induced fluorescence  is collected by a high numerical aperture lens. The single-shot fidelity of the detection is $98.6\pm0.2$\% and is presently limited by the dark count noise of the detector.
The simplicity of this method opens new perspectives in view of applications to
quantum manipulations of neutral atoms.

\end{abstract}


\maketitle

Cold and trapped neutral atoms and ions are model
systems to explore quantum computation~\cite{Nielsen,Blatt08,Ladd10}
and quantum simulation of many-body systems~\cite{Bloch08,Gerritsma08}.
They are also the basis of a variety of entangled states that open new and exciting perspectives for quantum metrology~\cite{Roos06}. For all these applications one requires
in particular
the control of the internal states on which quantum information is encoded. 
This can be done by using
laser manipulation techniques.
Another important requirement is the ability to read the internal state of the atoms
or ions with a high fidelity and a
minimal loss probability after performing a quantum manipulation.

A widely used method for state detection relies on fluorescence measurement~\cite{Wineland80}. To do so,
one identifies a bright and a dark state, which can be e.g. two hyperfine ground states,
separated by several GHz typically.
The bright state is coupled to an excited state by a closed optical transition.
The signature of the bright state  population is the emission of fluorescence light by the atom or ion when
it is illuminated by a probe laser tuned to this transition.
The signature of the dark state population, on the contrary, is the absence of fluorescence
due to the large hyperfine splitting. As this method relies on photon scattering, the energy of the probed atom or ion
increases with the number of recoils. In the case of ions, traps depths of several thousands of Kelvins are typical, leading to a very efficient state detection with a negligible loss probability.
There, detection fidelities as high as 99.99\% have been reported~\cite{Myerson08}.

Neutral atoms are also considered as forefront candidates for 
sophisticated quantum operations as one can handily control their interactions~\cite{Mandel03,Saffman10,Wilk10,Isenhower10}. They also provide 
built-in scalability when placed in optical lattices~\cite{Ladd10}. 
However, the detection  technique mentioned above, when applied to neutral atoms, 
is  hampered by the small trap depth, typically lower than a few milliKelvins.
The heating induced by the probe laser leads more easily to the loss of the atom
before one can collect enough photons to decide in which state the atom is.
Many groups have therefore implemented a  so-called  ``push-out" measurement based
on the state-selective loss of
the atom when it is illuminated by a resonant laser~\cite{Kuhr03,Yavuz06,Jones07}.
Although this technique has been proved to be
efficient and quantum projection limited~\cite{Jones07}, it
does not
discriminate between detection induced losses from any other
unwanted losses.
One is therefore in demand of a detection scheme that is not based on the atom loss.

One way to implement lossless and yet efficient detection is to place the trapped 
atom in an optical cavity.
Thanks to the Purcell effect, the fluorescence rate is enhanced in the cavity mode such that enough
fluorescence photons can now be collected without losing the atom.
Recently, two experiments demonstrated the state
selective detection of a single atom using an optical cavity with reported
fidelities larger than $99.4$\%~\cite{Bochmann10,Gehr10}.

In another approach one can simply make use of a lens with a high numerical aperture
to collect the fluorescence emitted by an atom trapped in an optical dipole trap.
In this Letter, we follow this route and demonstrate the single-shot detection of
the internal state of a rubidium 87 atom trapped in an optical tweezer. The fidelity of this state selective detection method
is $98.6$\% in $1.5$~ms and the probability to lose the atom during the detection is less than $2$\%.

\begin{figure}
\includegraphics[width=8cm]{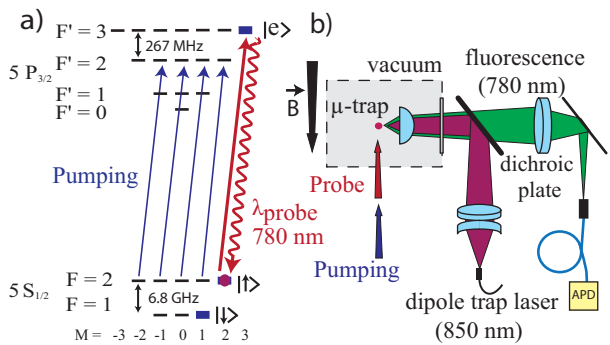}
\caption{(Color online)\textit{a}) Energy diagram of the D2 line of $^{87}$Rb.
\textit{b}) Experimental set-up. An aspheric lens focuses a $850$~nm laser
beam down to a spot with a $1.1~\mu$m waist ($\mu$-trap). Resonant probe light
induces the fluorescence of a single atom loaded in the microscopic trap, which
we detect on a fibered avalanche photodiode (APD).}
\label{Rb_Setup}
\end{figure}

The bright state used in our experiment is the hyperfine Zeeman state
$|\!\uparrow\rangle=|5S_{1/2},F=2,M=2\rangle$ (see Fig.~\ref{Rb_Setup}a). It is coupled to the excited state
$|e\rangle=|5P_{3/2},F'=3,M=3\rangle$ by a closed transition at $\lambda=780$~nm.
The dark state can be any Zeeman state of the $(5S_{1/2},F=1)$ manifold,
including $|\!\downarrow\rangle=|5S_{1/2},F=1,M=1\rangle$ \footnote{The states
$|\!\downarrow\rangle$ and $|\!\uparrow\rangle$ are commonly used as qubit states
and can be manipulated by micro-waves~\cite{Kuhr03} or Raman lasers~\cite{Yavuz06}.}.
It is separated from the bright state by  $\sim 6.835$~GHz.
Let us estimate the feasibility of the state detection, using probe light tuned
to the transition between $|\!\uparrow\rangle$ and $|e\rangle$
(the line width is $\Gamma/2\pi=6$~MHz and the saturation intensity is
$I_{\rm sat}=1.67$~mW.cm$^{-2}$). To do so, we consider an atom prepared at the bottom of the dipole trap in state $|\!\uparrow\rangle$ and
we estimate the number of absorption-spontaneous emission cycles that elevate the energy of
the atom by an amount equal to the trap depth $U$. For $U/k_{\rm B}=2$~mK (typical value for our experiment; $k_{\rm B}$ is the Boltzmann constant),
this number is on the order of $U/2E_{\rm r}\approx 5000$ ($E_{\rm r}=\frac{\hbar^2 k^2}{2m}$ is
the recoil energy induced by a photon with a wave vector $k=2\pi/\lambda$, $m$ is the mass of the atom).
This number puts constraints on the probe light parameters in order to detect the atom without losing it.
Taking for the saturation parameter $s=I/I_{\rm sat}=0.1$ and for the probe duration $\Delta t=1$~ms
yields a number of scattered photons $\frac{\Gamma}{2}\frac{s}{1+s}\Delta t\sim 2000$ during the probe pulse, below the 5000 photons calculated above. Using an imaging system with a detection efficiency of $0.6$\%~\cite{Darquie05}, one thus expects to detect $\sim11$ fluorescence photons in $1$~ms. As our noise level is well below $1$
photon during this time, the bright state $|\!\uparrow\rangle$ should be identified unambiguously. Based on this estimation we have implemented this method on a single atom.

Our experimental setup, represented in Fig.~\ref{Rb_Setup}b), has been described
in detail elsewhere~\cite{Sortais2007}. A single atom is trapped in
an optical tweezer produced by a linearly polarized $850$~nm laser. The optical tweezer is obtained by sharply focusing the laser
through an aspheric lens with a numerical aperture of 0.5. The atom is captured from an optical
molasses and
due to the small trapping volume only one atom is trapped
at a time~\cite{Sortais2007}. The temperature of the atom, measured by a
release-and-recapture technique~\cite{Tuchendler08}, is $35$~$\mu$K in a
2.2~mK deep trap.  This
value  ensures that the atom is close to the bottom of the trap.
We collect the fluorescence light emitted by the atom using the same aspheric lens.
This light is detected by a fibre coupled avalanche photodiode operating in a
single photon counting regime. At the beginning and at the end of each sequence
(see below), we test the presence of the atom in the trap by collecting the
fluorescence light induced by the molasses beams.
To detect the internal state of the atom, we send a unidirectional
$\sigma^{+}$-polarized probe laser propagating along the quantization axis
and tuned to the transition from
$|\!\uparrow\rangle$ to $|e\rangle$. The quantization axis is set
by applying a magnetic  bias field of $1$~G.

The experimental sequence begins with a single atom in a $2.2$~mK deep trap. After
switching off the molasses lasers we start the preparation phase in the bright state.
We  decrease adiabatically the trap
depth down to $0.24$~mK  and turn on the bias field. The state
preparation is achieved through optical pumping
by sending a $500$~$\mu$s pulse of pumping light tuned
to the transition $(5S_{1/2},F=2)$ to $(5P_{3/2},F'=2)$ superimposed to
repumping light tuned to the transition $(5S_{1/2},F=1)$ to $(5P_{3/2},F'=2)$.
Both laser beams propagate along the quantization axis
and are $\sigma^{+}$-polarized.
We measure the fidelity of the preparation in $(5S_{1/2},F=2)$ by
using the ``push-out" technique mentioned in the introduction.
We find a hyperfine state preparation efficiency of $99.97$\%, obtained by recapturing
2  atoms after $6000$ cycles. In order to estimate the efficiency of
the preparation in, more specifically, the $M=2$ Zeeman state, we analyze
the heating rate induced by the preparation light based on the following fact:
an atom well prepared
in state $|\!\uparrow\rangle$ would not scatter photons when illuminated by the
preparation light beams and would thus remain in the dipole trap unheated.
Measuring the actual heating rate thus yields the probability for the atom
to be in other Zeeman states of the $F=2$ manifold.  We deduce
a preparation efficiency in $|\!\uparrow\rangle$
of $99.6$\%. To conclude on the state preparation,
we can alternatively prepare the atom in the $F = 1$ manifold  (with no control over the
Zeeman states) by illuminating the atom with the pumping light only.

We now turn to the state selective detection phase. At the end of the preparation
stage we ramp up the trap depth to a value $U$ and send the state detection
probe light during a time $\Delta t$. The probe
is tuned on resonance with light-shifted transition of the atom at the bottom the trap.
The saturation parameter of the
(unidirectional) probe is chosen sufficiently low that the effect of
the potential due to
the radiation pressure force is negligible on the dipole trap depth.
During this probing sequence, we count the number of photons detected on the
avalanche photodiode. Depending on the value of the trap depth, we
adjust the duration of the probe pulse such that it induces less than 2\% atom loss  
whilst maximizing the number of detected photons. 
Note that in addition to the probe induced losses, we measure  
an atom loss probability of  1\% intrinsic to our set up, due to errors when testing
for the presence of the atom at the beginning and at the end of the sequence ($0.6$\%)
and to the vacuum limited lifetime $\tau=23$~s of the single atom in the
dipole trap ($0.4$\%). In the results presented below
we post-select the experiments where the atom is present at the end of the
sequence.

\begin{figure}
\includegraphics[width=8cm]{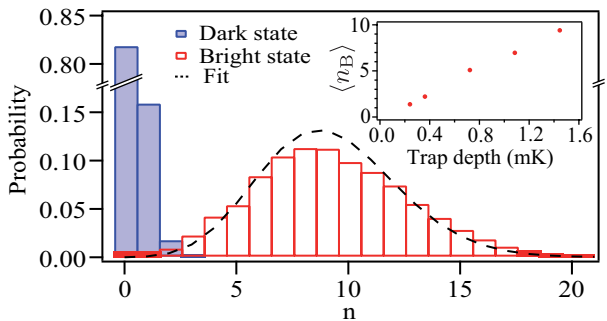}
\caption{(Color online) Histogram of the probability to detect a
number of photons $n$ when the atom is initially prepared in the dark state
(blue bars) or in the bright state
(red bars). The trap depth is $U/k_{\rm B}=1.4~$mK, the duration of the probe
$\Delta t=1.5~$ms, and the saturation $s=6.1\times10^{-2}$.
The histograms correspond to an average of $9700$ sequences.
The black dashed line is a Poissonian fit to the data.
(Inset) Mean number of detected photons $\langle n_B\rangle$ versus the 
trap depth $U$ for an atom prepared in the bright state.}
\label{distribution}
\end{figure}

The result of the experiment for
a value of the trap depth of $U=1.4$~mK
is shown in Fig.~\ref{distribution}. This figure represents two
histograms of the number $n$ of photons detected during the probing period,
for an atom prepared respectively in the dark state (distribution $P_{\rm D}(n)$) and
in the bright state $|\!\uparrow\rangle$
(distribution $P_{\rm B}(n)$). In  the case of the dark state
the distribution is very close to Poissonian with a mean value 
$\langle n_D\rangle=0.2$ photon. We performed the same experiment
when no atom is present and found no deviation with respect to
the case where the atom is prepared in the dark state.
The dark state signal thus comes from the background only, i.e. it corresponds to a dark count rate 
of the avalanche photodiode of $130$ counts/sec. In the case of the preparation in  $|\!\uparrow\rangle$, the histogram
shows a mean number of detected photons $\langle n_B\rangle=9.2$
much larger than for the dark state, as expected. The distribution is also nearly
Poissonian.

We also varied the trap depth and optimized
the duration and the saturation parameter of the probe, as explained above.
For the set of trap depths $U/k_{\rm B}=(0.24; 0.36; 0.7; 1.1; 1.4)$~mK
the probe duration was respectively
$\Delta t =(0.7; 0.75; 1; 1.25; 1.5)$~ms and the saturation parameter was
$s=(1.1; 1.9; 3.7; 4.9; 6.1)\times 10^{-2}$. The inset in
Fig.~\ref{distribution} summarizes the results on the mean number of detected
photons $\langle n_B\rangle$ versus the trap depth $U$ for an atom prepared in the bright state.
The linear dependence indicates that the average number of photons scattered
by the trapped atom varies proportionally to the trap depth, as assumed qualitatively at the beginning of this Letter.
As a side result, we compare $\langle n_B\rangle$
to the number of photons scattered by the atom during the probe pulse, $\frac{\Gamma}{2}\frac{s}{1+s} \Delta t $.
This yields a collection efficiency  of $\approx0.6\%$ for our imaging system, in good agreement with an
independent estimate of the solid angle of
the aspheric lens ($7$\%), the transmission of the optics including the fiber coupling ($20$\%) and
the quantum efficiency of the detector ($50$\%).

In order to characterize the performance of the state detection we use the
state readout fidelity ${\rm \cal F}$ defined in~\cite{Myerson08}:
\begin{equation}
\label{fid}
{\rm \cal F}=1-\frac{1}{2}(\epsilon_{\rm B}+\epsilon_{\rm D})
\end{equation}
where $\epsilon_{\rm B}$ is the fraction of experiments in which an atom
prepared in the bright state is detected to be dark and, conversely,
$\epsilon_{\rm D}$ is the fraction of experiments where an atom prepared
in the dark state is found to be bright. To calculate these quantities
we define a threshold $n_c$ on the number of detected photons.
We consider that an experiment where more (resp. less) than $n_c$ photons are detected during the
probe pulse corresponds to an atom prepared in the bright (resp. dark) state.
We  calculate the
errors $\epsilon_{\rm B}$ and $\epsilon_{\rm D}$ using
\begin{equation}
\label{erreurs}
\epsilon_{\rm B}=\sum_{n=0}^{n_c}P_{\rm B}(n)~~
\text{and} ~~\epsilon_{\rm D}=\sum_{n=n_c+1}^{\infty}P_{\rm D}(n)\ .
\end{equation}

\begin{figure}
\includegraphics[width=8cm]{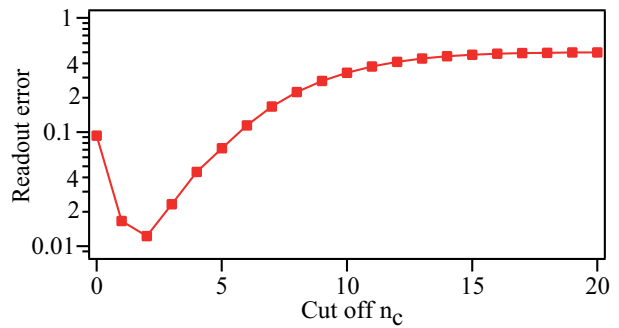}
\caption{(Color online) Readout error $\epsilon = 1-{\rm \cal F}$ versus
 the threshold on the number of detected photons, $n_{c}$, for the data of Fig.~\ref{distribution}.}
\label{Cutoff}
\end{figure}

Figure~\ref{Cutoff} shows the readout error
$\epsilon=\frac{1}{2}(\epsilon_{\rm B}+\epsilon_{\rm D})$
versus the threshold $n_c$, for the same set of data as in Fig.~\ref{distribution}.
From the data shown in Fig.~\ref{Cutoff}, we extract a minimal readout error of
$1.2$\% (obtained for $n_c=2$).  This is equivalent to a state detection fidelity ${\rm \cal F}=98.8$\%.
We repeated the same experiment $6$ times over several days and found an average
fidelity ${\rm \cal F}=98.6\pm0.2\%$ (the error bar is
statistical).

\begin{figure}
\includegraphics[width=8cm]{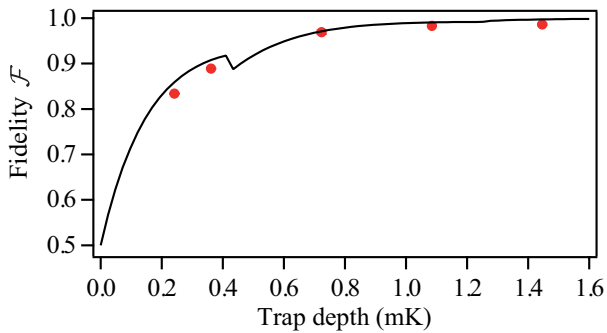}
\caption{(Color online)  State detection fidelity versus the trap depth.
The black line is a model explained in the text.}
\label{Fidelity}
\end{figure}

Large values of the trap depth allow us to increase the probe durations to detect more photons. As
the background rate (dark count rate of the detector)
remains constant, the fidelity increases, as shown in Fig.~\ref{Fidelity}. We compare these data
to a model using Poissonian distributions for $P_{\rm D }(n)$ and $P_{\rm B }(n)$ with mean values
$\langle n_{D}\rangle$ and $\langle n_{B}\rangle$ that depend on the trap depth as discussed above (see inset in Fig.~\ref{distribution}).
The optimum threshold $n_c$ is calculated for
each value of the trap depth. Not surprisingly, we find good agreement between our data and the 
model~\footnote{The discontinuities in the model are a direct consequence of $n_c$ varying by 
integer values when the trap depth increases.}.

Finally, we discuss the factors that limit our state detection fidelity to $98.6$\% and explore the possibilities for improvement in the future. The main contribution to the error budget (see table~\ref{table_errorbudget}) comes from the dark counts of our avalanche photodiode and is $\approx1$\%. Using commercially available photodiodes with a lower dark count rate of $25~{\rm s}^{-1}$~\cite{Bochmann10} would readily bring this error contribution down to $0.3$\%. A small contribution to the error budget comes from the above mentioned imperfect state preparation in $|\!\uparrow\rangle$ ($0.03$\%). Off-resonant Raman transitions induced by the dipole trap light after the preparation phase also contribute for $\approx0.1$\% as they mimic a bad state preparation by coupling the $F=1$ and $F=2$ levels~\cite{Cline94}. As this contribution scales approximately as $\Delta^{-4}$ ($\Delta$ is the trap laser frequency detuning with respect to the fluorescence transitions), we estimate that using a trapping laser with a larger wavelength while maintaining the same trap depth would efficiently reduce this error. The last contribution, which is presently $0.27$\%, comes from the small number of detected photons, which leads to a small value for $\langle n_B\rangle$ and hence to a non negligible value for $P_{\rm B }(n=0)$. This error will be hard to 
reduce significantly as $P_{\rm B }(n=0)=e^{-\langle n_B\rangle}$  varies slowly for large values of $\langle n_B\rangle$.

\begin{table}[h]
  \centering
  \begin{tabular}{|c|c|}
  \hline
  Source of error & Contribution \\
  \hline
  Detector dark counts & 1\% \\
  Detection inefficiency & 0.27\% \\
  Raman transitions & 0.1\% \\
  Imperfect preparation & 0.03\% \\
  \hline
  Total error & 1.4\% \\
    \hline
\end{tabular}
\caption{Error budget of our lossless state detection.}\label{table_errorbudget}
\end{table}

In conclusion, we have demonstrated  a lossless internal state readout
of a single atom trapped in an optical tweezer. This method is
based on the collection of the probe induced fluorescence
using a simple imaging optics.
The fidelity of the state detection is presently $98.6$\% in single shot, with room for technical improvements in the future. 
Combined with our ability to efficiently control the internal states of single atoms~\cite{Jones07,Wilk10}, this non-destructive
state detection completes our toolbox for quantum engineering. We therefore 
believe that the detection presented in this Letter will be of great interest for future applications involving quantum measurements. Furthermore,  
the absence of  atom loss will prevent the reloading of the atom after each measurement, 
thus improving the duty cycle of the experiments.
It will also avoid post-detection corrections
when performing quantum operations on a set of neutral atom qubits.

\begin{acknowledgments}

{\it Note -} During the preparation of this manuscript, we have learned of the
existence of a related work~\cite{Gibbons10}.

We acknowledge support from the European Union through the ERC Starting Grant ARENA.
A.~Fuhrmanek acknowledges partial support from the DAAD Doktorandenstipendium.
\end{acknowledgments}

\end{document}